\documentclass[]{an}
\usepackage{graphicx,times}
\usepackage{bm}
\graphicspath{{./fig/}{./png/}}

\newcommand{\EQ}{\begin{equation}}
\newcommand{\EN}{\end{equation}}
\newcommand{\EQA}{\begin{eqnarray}}
\newcommand{\ENA}{\end{eqnarray}}

\newcommand{\Eq}[1]{Eq.~(\ref{#1})}
\newcommand{\Eqs}[2]{Eqs.~(\ref{#1}) and~(\ref{#2})}

\newcommand{\eqs}[2]{(\ref{#1}) and~(\ref{#2})}

\newcommand{\Sec}[1]{Sect.~\ref{#1}}

\newcommand{\Fig}[1]{Fig.~\ref{#1}}

\newcommand{\Tab}[1]{Table~\ref{#1}}

\newcommand{\meanrho}{\overline{\rho}}

{}
{}
\newcommand{\meanFFFF}{\overline{\mbox{\boldmath ${\cal F}$}}{}}{}

\newcommand{\meanSSSS}{\overline{\mbox{\boldmath ${\mathsf S}$}} {}}

{}
{}
{}
{}
{}
{}
{}
\newcommand{\meanAA}{\overline{\mbox{\boldmath $A$}}{}}{}
\newcommand{\meanBB}{\overline{\mbox{\boldmath $B$}}{}}{}
{}
\newcommand{\meanFF}{\overline{\mbox{\boldmath $F$}}{}}{}
{}
{}
{}
{}
{}
\newcommand{\meanJJ}{\overline{\mbox{\boldmath $J$}}{}}{}
\newcommand{\meanUU}{\overline{\bm{U}}}

{}
{}
{}
\newcommand{\meanA}{\overline{A}}
\newcommand{\meanB}{\overline{B}}

\newcommand{\meanU}{\overline{U}}

{}

{}
{}

\newcommand{\uu}{\mbox{\boldmath $u$} {}}
\newcommand{\UU}{\mbox{\boldmath $U$} {}}

\newcommand{\bb}{\mbox{\boldmath $b$} {}}
\newcommand{\BB}{\mbox{\boldmath $B$} {}}

\newcommand{\grav}{\mbox{\boldmath $g$} {}}
\newcommand{\nab}{\mbox{\boldmath $\nabla$} {}}

\newcommand{\ssigma}{\mbox{\boldmath $\sigma$} {}}
\newcommand{\SSSS}{\mbox{\boldmath ${\sf S}$} {}}

\def\Rm{R_{\rm m}}

\def\Rey{\mbox{\rm Re}}

\def\csz{c_{\rm s0}}
\def\cs{c_{\rm s}}

\def\ptz{p_{\rm t0}}

\def\lf{l_{\rm f}}
\def\kf{k_{\rm f}}

\def\urms{u_{\rm rms}}

\def\nut{\nu_{\rm t}}
\def\etat{\eta_{\rm t}}

\def\Beq{B_{\rm eq}}

\def\half{{\textstyle{1\over2}}}

\def\onethird{{\textstyle{1\over3}}}

\newcommand{\G}{\,{\rm G}}

\newcommand{\Mm}{\,{\rm Mm}}

\newcommand{\yapj}[3]{ #1, {ApJ,} {#2}, #3}

\newcommand{\yan}[3]{ #1, {Astron.\ Nachr.,} {#2}, #3}

\newcommand{\yana}[3]{ #1, {A\&A,} {#2}, #3}

\newcommand{\ygafd}[3]{ #1, {Geophys.\ Astrophys.\ Fluid Dyn.,} {#2}, #3}

\newcommand{\ypf}[3]{ #1, {Phys.\ Fluids,} {#2}, #3}

\newcommand{\ysovl}[3]{ #1, {Sov.\ Astron.\ Lett.,} {#2}, #3}
\newcommand{\yjetp}[3]{ #1, {Sov.\ Phys.\ JETP,} {#2}, #3}

\newcommand{\ymn}[3]{ #1, {MNRAS,} {#2}, #3}
\newcommand{\ynat}[3]{ #1, {Nature,} {#2}, #3}

\newcommand{\ysci}[3]{ #1, {Science,} {#2}, #3}
\newcommand{\ysph}[3]{ #1, {Solar Phys.,} {#2}, #3}
\newcommand{\ypr}[3]{ #1, {Phys.\ Rev.,} {#2}, #3}
\newcommand{\ypre}[3]{ #1, {Phys.\ Rev.\ E,} {#2}, #3}

\newcommand{\ybook}[3]{ #1, {#2} (#3)}

\title{Large-scale magnetic flux concentrations from turbulent stresses}
\titlerunning{Large-scale magnetic flux concentrations}
\authorrunning{A. Brandenburg et al.}
\author{Axel Brandenburg\inst{1,2}, Nathan Kleeorin\inst{3}, Igor Rogachevskii\inst{3}}
\institute{NORDITA, AlbaNova University Center, Roslagstullsbacken 23,
SE 10691 Stockholm, Sweden
\and
Department of Astronomy, AlbaNova University Center,
Stockholm University, SE 10691 Stockholm, Sweden
\and
Department of Mechanical
Engineering, The Ben-Gurion University of the Negev, POB 653,
Beer-Sheva 84105, Israel
}
\date{\today,~ $ $Revision: 1.113 $ $}
\abstract{%
In this study we provide the first numerical demonstration of the effects
of turbulence on the mean Lorentz force and
the resulting formation of large-scale magnetic structures.
Using three-dimensional direct numerical simulations (DNS)
of forced turbulence
we show that an imposed mean magnetic field leads
to a decrease of the turbulent hydromagnetic pressure and tension.
This phenomenon is quantified by determining the relevant functions that
relate the sum of the turbulent Reynolds and Maxwell stresses with the
Maxwell stress of the mean magnetic field.
Using such a parameterization, we show by means of two-dimensional
and three-dimensional
mean-field numerical modelling that an isentropic density stratified
layer becomes unstable in the presence of a uniform imposed magnetic field.
This large-scale instability results in the formation of
loop-like magnetic structures which are concentrated at the top
of the stratified layer.
In three dimensions these structures resemble the appearance
of bipolar magnetic regions in the Sun.
The results of DNS and mean-field numerical modelling are
in good agreement with theoretical predictions.
We discuss our model in the context of a distributed solar dynamo
where active regions and sunspots might be rather shallow phenomena.}

\begin{document}
\maketitle

\section{Introduction}

Turbulence effects generally refer to the occurrence of
corre\-la\-tions between velocity, temperature,
and/or magnetic fields at small scales.
A typical example is turbulent viscosity, which results from the spatial
exchange of turbulent eddies characterized by velocity correlations.
This leads to the dissipation of energy at small scales.

However, there is also the possibility of additional
(e.g.\ non-diffusive) turbulence effects, as
is perhaps best known in mean-field electrodynamics and dynamo theory.
Here one models the effects of the mean electromotive force, i.e.\
the turbulence effects of velocity and magnetic field fluctuations,
on the evolution of the mean field.
This can lead to the occurrence of the famous $\alpha$ effect, in addition
to turbulent magnetic diffusion, turbulent diamagnetic velocity, etc.
(Moffatt 1978; Krause \& R\"adler 1980).
Another example is the $\Lambda$ effect in rotating anisotropic
hydrodynamic turbulence, which can lead to the occurrence of
differential rotation in cosmic bodies such as the Sun
(R\"udiger 1980, 1989; R\"udiger \& Hollerbach 2004).
In that case the relevant correlations come from the mean Reynolds
stress tensor and its dependence on the local angular velocity.

A related example is the combined Reynolds
and Max\-well turbulent stress tensor
and its dependence on the mean magnetic field.
The first analytic calculations of the dependence of the turbulent
Reynolds stress on the mean magnetic field
in the framework of the first-order smoothing approximation
were performed by R\"adler (1974) and R\"udiger (1976).
Later, also the combined effects of the turbulent Reynolds
and Maxwell stress tensors were considered (Kleeorin et al.\ 1989,
1990; R\"udiger \& Kichatinov 1990). It was noticed that this can lead to
a local reduction of the total turbulent pressure and hence to
the possibility of self-induced concentrations of large-scale
magnetic fields (Kleeorin et al.\ 1989, 1990, 1996;
Kleeorin \& Rogachevskii 1994; Rogachevskii \& Kleeorin 2007).
Such a process may play an important role in the formation of sunspots and
active regions in the Sun. It may be complementary to
the magnetically induced suppression
of the turbulent energy flux, which would lead to further cooling and
hence a further concentration of the structures (Kitchatinov \& Mazur
2000).

The current leading explanation for the formation of sun\-spots is related to
the emergence of deeply rooted magnetic flux tubes (Parker 1955, 1982, 1984).
Such flux tubes are generally believed to be produced and `stored' near
the bottom of the convection zone (Spiegel \& Weiss 1980).
The storage of magnetic fields and the formation of flux tubes in
the overshoot layer near the bottom of the solar convective zone was
investigated in a number of publications (see, e.g., Spruit 1981; Spruit
\& van Ballegooijen 1982; Sch\"{u}ssler et al.\ 1994; Moreno-Insertis
et al.\ 1996; Tobias et al.\ 2001; Tobias \& Hughes 2004).
However, in order that the tubes retain their basic east--west orientation
during their ascent over many pressure scale heights, the magnetic field
must be strong enough (Choudhuri \& D`Silva 1990) and is estimated to
be around $10^5\G$ at the bottom of the convection zone
(D`Silva \& Choudhuri 1993).
Such fields would be up to a hundred times stronger than the equipartition
value, which is one of several other arguments that have led to the idea
that flux emergence of dynamo-generated fields might instead be a
shallow phenomenon (Brandenburg 2005; Schatten 2009).
Such a scenario appears also compatible with solar subsurface flows,
as inferred from local helioseismology (Zhao et al.\ 2001; Kosovichev 2002).
In particular, Zhao et al.\ (2004) and Hindman et al.\ (2009)
find the presence of converging
flows around active regions at radii as large as 100--200\Mm.
It appears that these convergent flows might actually be the
source of the formation of active regions and perhaps sunspots
rather than a consequence (Parker 1979a; Hurlburt \& Rucklidge 2000).
Of course, in the immediate proximity of individual spots one observes
outgoing flows. Those are probably superficial, less than 1--2\Mm\ deep,
and appear to be due to the dynamical effects of magnetoconvection
in an inclined magnetic field of the penumbra (e.g., Thomas et al.\ 2002;
Heinemann et al.\ 2007; Rempel et al.\ 2009; Kitiashvili et al.\ 2009).

The goal of this paper is to investigate the effects of turbulence
on the mean Lorentz force by means of direct numerical simulations (DNS)
for forced turbulence and to study the instability
of a uniform large-scale magnetic field in an adiabatically
stratified layer by means of mean-field numerical modelling
based on parameterizations both of analytic formulae by
Rogachevskii \& Kleeorin (2007) and the results of DNS for forced turbulence.
In order to study the essence of the effect, we make several
simplifications by neglecting the energy equation, i.e.\ the specific
entropy is assumed to be strictly constant in space and time.
In the mean-field numerical modelling we
neglect the suppression of turbulent magnetic diffusivity and
turbulent viscosity, and omit correlations with density fluctuations.
Nevertheless, the mean density is allowed to evolve fully
self-consistently according to the usual continuity equation.

\section{Turbulence effects on mean Lorentz force}

Throughout this paper we adopt units for the magnetic field where the
vacuum permeability is equal to unity, i.e.\ the magnetic pressure
is given by $\half\BB^2$.

\subsection{General considerations}

We use the equations of mean-field magnetohydrodynamics (MHD).
These equations are obtained by averaging the original MHD
equations over small-scale fluctuations.
This technique is best known in the case of the induction equation
(Moffatt 1978; Krause \& R\"adler 1980).
In this study we are mainly interested in effects of turbulence
on the mean Lorentz force. Let us consider the momentum equation,
\EQ
{\partial\over\partial t}\rho\, \UU=-{\partial\over\partial x_j}\Pi_{ij},
\EN
where
\EQ
\Pi_{ij}=\rho \, U_iU_j+\delta_{ij}\left(p+\half\BB^2\right)-B_iB_j-\sigma_{ij}
\label{Piorig}
\EN
is the momentum stress tensor, $\UU$ and $\BB$ are the velocity and
magnetic fields, $p$ and $\rho$ are the fluid pressure and density,
$\delta_{ij}$ is the unit Kronecker tensor,
$\sigma_{ij}=2\rho\nu{\sf S}_{ij}$ is the viscous stress tensor, with
\EQ
{\sf S}_{ij}=\half(U_{i,j}+U_{j,i})-\onethird\delta_{ij}\nab\cdot\UU
\label{strain}
\EN
being the traceless rate of the strain tensor,
and $\nu$ is the kinematic viscosity.
Ignoring the turbulent correlations with density fluctuations
for low-Mach number turbulence, the averaged momentum
equation is
\EQ
{\partial\over\partial t} \meanrho \, \meanUU=
-{\partial\over\partial x_j}\overline{\Pi}_{ij},
\EN
where $\overline\Pi_{ij}=\overline\Pi_{ij}^{\rm m}+\overline\Pi_{ij}^{\rm f}$
is the mean momentum stress tensor split into contributions resulting
entirely from the mean field (indicated by superscript m) and those of
the fluctuating field (indicated by superscript f).
The tensor $\overline\Pi_{ij}^{\rm m}$ has the same form as \Eq{Piorig},
but all quantities have now attained an overbar, i.e.\
\EQ
\overline\Pi_{ij}^{\rm m}=\meanrho \, \meanU_i\meanU_j
+\delta_{ij}\left(\overline{p}+\half\meanBB^2\right)
-\meanB_i\meanB_j-\overline\sigma_{ij}.
\EN
We emphasize here that $\overline{p}$ is just the mean gas pressure and
$\overline\sigma_{ij}$ is the average of the microscopic viscous stress
tensor, $\sigma_{ij}$.
The contribution from the fluctuating fields, in turn, is split into
contributions that are independent of the mean fields (and hence isotropic
and proportional to $\delta_{ij}$) and
contributions which depend on the mean fields,
\EQ
\overline\Pi_{ij}^{\rm f}=\ptz\delta_{ij}-\sigma_{ij}^{\rm eff}.
\EN
Here, $\ptz$ is the turbulent pressure in the absence of a mean magnetic
field and $\sigma_{ij}^{\rm eff}=\sigma_{ij}^{\rm K}+\sigma_{ij}^{\rm M}$
quantifies the turbulent viscosity,
$\sigma_{ij}^{\rm K}=2\meanrho\nu_t \overline{\sf S}_{ij}$,
and the additional effects of the mean magnetic field
on the effective stress tensor.
Both terms, $\ptz\delta_{ij}-\sigma_{ij}^{\rm M}$,
result from the fluctuations $\uu=\UU-\meanUU$ and
$\bb=\BB-\meanBB$ of velocity and magnetic fields, and are given by
\EQ
\ptz\delta_{ij}-\sigma_{ij}^{\rm M}=
\meanrho \, \overline{u_iu_j}+\half\delta_{ij}\overline{\bb^2}-\overline{b_ib_j}.
\label{stress0}
\EN
In the absence of a mean magnetic field, the turbulent background pressure is
\EQ
\ptz={\textstyle{1\over6}}\overline{\bb_0^2}
+{\textstyle{1\over3}}\rho\overline{\uu_0^2}
\label{pt}
\EN
where the subscripts 0 on $\bb^2$ and $\uu^2$ indicate values in the
absence of the mean magnetic field.
Magnetic fluctuations $\bb$
are generated both by small-scale dynamo action and by
tangling of the mean magnetic field by velocity fluctuations.
On the other hand, the velocity fluctuations also have two contributions,
those that depend on the mean magnetic field and those that do not.

Following Rogachevskii \& Kleeorin (2007), the part of
the effective stress tensor that depends on the
mean magnetic field is parameterized as
\EQ
\sigma_{ij}^{\rm M}(\meanBB)=-q_{\rm s}\meanB_i\meanB_j+\half\delta_{ij}
q_{\rm p}\meanBB^2,
\label{stress1}
\EN
where $q_{\rm s}$ and $q_{\rm p}$ are functions of the mean field, $\meanBB$,
and the magnetic Reynolds number, ${\rm Rm}$.
\Eq{stress1} implies that the effective mean Lorentz force
that takes into account the effects of turbulence, can be written as:
\EQ
\meanrho \, \meanFFFF_{\rm M} = -\half\nab[(1-q_{\rm p})\meanBB^2]
+ \meanBB \cdot \nab\left[(1-q_{\rm s})\meanBB\right].
\label{efforce}
\EN
The detailed analytic expressions for $q_{\rm s}(\meanB)$ and $q_{\rm p}(\meanB)$
have been given by Rogachevskii \& Kleeorin (2007). The asymptotic formulae
for the nonlinear functions, $q_{\rm p}(\meanB)$ and $q_{\rm s}(\meanB)$, are given below.
For this purpose we define $\beta \equiv \meanB/B_{\rm eq}$, where
$B_{\rm eq}=(\rho\overline{\uu_0^2})^{1/2}$ is the equipartition field strength.

For very weak mean magnetic fields, $4\beta \ll {\rm Rm}^{-1/4} $,
$\, q_{\rm p}$ and $q_{\rm s}$ are approximately constant and given by
\begin{eqnarray*}
q_{\rm p}(\beta) &=& {4 \over 45} \, \big(1 + 9\ln {\rm Rm}\big)\;,
\\
q_{\rm s}(\beta) &=& {2 \over 15} \, \big(1 + 4\ln {\rm Rm}\big) \;,
\end{eqnarray*}
for $ {\rm Rm}^{-1/4} \ll 4\beta \ll 1$ we have
\begin{eqnarray*}
q_{\rm p}(\beta) &=& {16 \over 25} \, [1 + 5|\ln (4 \beta)| + 32
\, \beta^{2}] \;,
\\
q_{\rm s}(\beta) &=& {32 \over 15} \, \biggl[|\ln (4 \beta)| +
{1 \over 30} + 12  \beta^{2} \biggr] \;,
\end{eqnarray*}
and for strong fields, $4\beta \gg 1 $, we have
\begin{eqnarray*}
q_{\rm p}(\beta) &=& 1/6\beta^2 \;,
\quad q_{\rm s}(\beta) = \pi/48\beta^3 \; .
\end{eqnarray*}
(Rogachevskii \& Kleeorin 2007).

In \Sec{TurbSim} we present DNS evidence that the functions
$q_{\rm p}(\meanB)$ and $q_{\rm s}(\meanB)$ are
positive, indicating the possibility of a reduction of the effective
Lorentz force, i.e., a decrease of the effective magnetic pressure
and magnetic tension in small-scale turbulence.

\subsection{DNS of turbulence effects on mean Lorentz force}
\label{TurbSim}

In order to study turbulence effects on the mean Lorentz force and to determine
the functions $q_{\rm p}(\meanB)$ and $q_{\rm s}(\meanB)$ from DNS, we
consider forced turbulence in a periodic three-dimensional domain in the
presence of an imposed uniform magnetic field, say $\meanBB_0=(\meanB_0,0,0)$.
We determine $q_{\rm s}$ and $q_{\rm p}$ from \Eqs{stress0}{stress1} for $i=j=x$,
\EQ
\ptz+\half(\overline{b_x^2}-\overline{b_y^2}-\overline{b_z^2})
-\meanrho \, \overline{u_x^2}=(\half q_{\rm p}-q_{\rm s})\meanB_0^2,
\label{stress2}
\EN
and $i=j=y$,
\EQ
\ptz+\half(\overline{b_y^2}-\overline{b_x^2}-\overline{b_z^2})
-\meanrho \, \overline{u_y^2}=\half q_{\rm p}\meanB_0^2,
\label{stress3}
\EN
where $\ptz$ is given by \Eq{pt}.
First, we determine $\ptz$ from a simulation with $\meanB_0=0$.
Then, we use \Eq{stress3} to determine $q_{\rm p}(\meanBB)$.
Finally, to determine $q_{\rm s}(\meanBB)$ we subtract \Eq{stress2} from
\Eq{stress3}, i.e.\
\EQ
(\overline{b_y^2}-\overline{b_x^2})
-\meanrho \, (\overline{u_y^2}-\overline{u_x^2})=q_{\rm s}\meanB_0^2.
\label{stress4}
\EN
In order to determine separately the effects of the mean field on the
turbulent Maxwell and Reynolds stresses we also compute their respective
contributions $q_{\rm p}^{\rm M}+q_{\rm p}^{\rm K}= q_{\rm p}$ and
$q_{\rm s}^{\rm M}+q_{\rm s}^{\rm K}= q_{\rm s}$.
In the current case where there is no small-scale dynamo action we have
\EQ
q_{\rm p}^{\rm M}
=(\overline{b_y^2}-\overline{b_x^2}-\overline{b_z^2})/\meanB_0^2,\quad
q_{\rm p}^{\rm  K}=2(\ptz-\meanrho \, \overline{u_y^2})/\meanB_0^2,\;\;
\label{stressMK}
\EN

\begin{figure}[t!]\begin{center}
\includegraphics[width=\columnwidth]{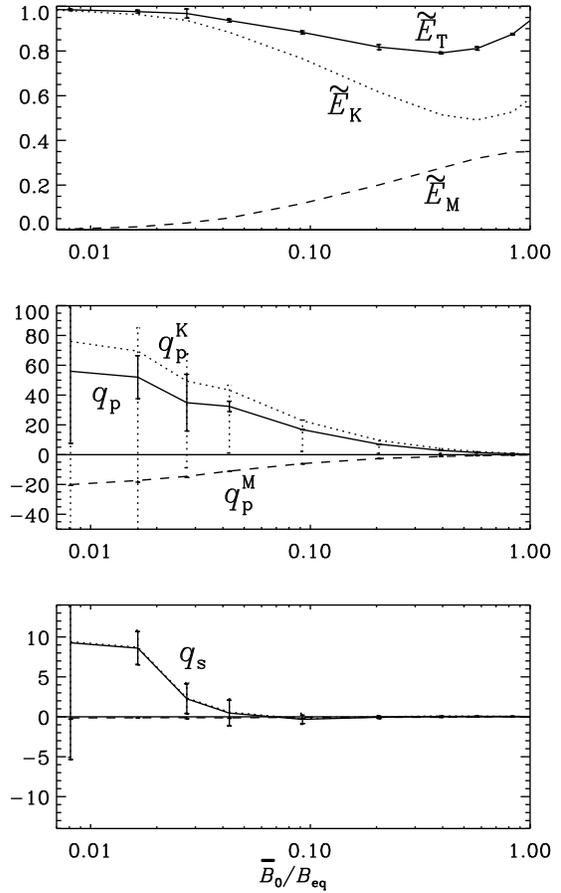}
\end{center}\caption[]{
$\meanB_0$ dependence of the
normalized turbulent energy $\tilde{E}_{\rm T}=E_{\rm T}/E_0$,
where $E_0$ is the value of $E_{\rm T}$ for $\meanB_0=0$,
together with the contributions from kinetic and magnetic energies,
$\tilde{E}_{\rm K}=E_{\rm K}/E_0$ and $\tilde{E}_{\rm M}=E_{\rm M}/E_0$,
(upper panel) as well as the coefficients $q_{\rm p}$,
$q_{\rm p}^{\rm K}$, $q_{\rm p}^{\rm M}$,
and $q_{\rm s}$ obtained from DNS
for $\Rey=180$ and $\Rm=45$ (second and third panels).
}\label{pP64_pq}\end{figure}

In \Fig{pP64_pq} we show the results from the simulations of forced
turbulence in the presence of an imposed magnetic field.
Here we have also plotted the total turbulent energy,
$E_{\rm T}=E_{\rm K}+E_{\rm M}$ where
\EQ
E_{\rm K}=\meanrho \, \overline{\uu^2}/2,\quad
E_{\rm M}=\overline{\bb^2}/2
\label{EK_EM_def}
\EN
are the energy densities of velocity and magnetic fluctuations,
respectively.
The simulations were performed with the {\sc Pencil Code}%
\footnote{{http://www.nordita.org/software/pencil-code/}},
which uses sixth-order explicit finite differences in space and third
order accurate time stepping method.
They are similar to the DNS of Haugen \& Brandenburg (2004).
The forcing function consists of plane non-polarized waves with
an average wavenumber $\kf=1.5\,k_1$, where $k_1=2\pi/L$ is the smallest
wavenumber that fits into a 3D periodic domain of size $L$.
We use an isothermal equation of state with constant sound speed $\cs$.
The forcing strength is arranged such that the turbulent rms velocity,
$\urms$, is below $\cs$.
In all our runs the maximum turbulent Mach number, $\urms/\cs$, is around 0.2,
so that compressibility effects are still weak.

The fluid Reynolds number based on the forcing wave number
$\kf=2\pi /\lf$, is $\Rey=\urms/\nu\kf\approx180$ and
the magnetic Prandtl number is $\nu/\eta=0.25$, so the magnetic Reynolds
number, $\Rm=\urms/\eta\kf\approx45$, is just small enough so that
no small-scale dynamo is excited.
[Following Haugen et al.\ (2004) and Schekochihin et al.\ (2005, 2007),
the critical value of $\Rm$ is between 70 and 80 for this value of the
magnetic Prandtl number]. Note that the magnetic Reynolds number,
${\rm Rm}$, used in Sect. 2.1 is based on forcing scale $\lf=2\pi/\kf$,
and thus ${\rm Rm} = 2 \pi \Rm$.
In all cases we normalize the field strength in terms of the equipartition
value, $\Beq=\rho_0^{1/2}\urms$. The number of mesh points is $64^3$.

As follows from our DNS study, both functions $q_{\rm p}(\meanB)$ and
$q_{\rm s}(\meanB)$ are positive and exceed unity for weak fields.
The error bars are obtained by calculating averages over each of three
equally long intervals of the full time series and taking the largest
deviation from the full averages.
For very small values of $B_0/\Beq$, the turbulent fluctuations
dominate over the effects from shredding the imposed field, which
increases the error bars.
Nevertheless, the results are in agreement with the prediction
of Rogachevskii \& Kleeorin (2007).
Note also that both theory and simulations suggest that $q_{\rm p}>2q_{\rm s}$.

\subsection{Physics of turbulence effects on Lorentz force}

The physics of the effect of turbulence on the large-scale Lorentz force is as follows.
The equation of state for isotropic turbulence is given by
\EQ
\ptz = \frac{1}{3} E_{\rm M} + \frac{2}{3} E_{\rm K} \;,
\label{eqstate}
\EN
[see, e.g., Landau \&  Lifshitz 1975, 1984, and  \Eq{pt}], where $\ptz$
is the total (hydrodynamic plus magnetic) turbulent pressure,
and $E_{\rm K}$ and $E_{\rm M}$ are defined by \Eq{EK_EM_def}.
The total energy density $E_{\rm T} = E_{\rm K} + E_{\rm M}$
of homogeneous turbulence
with a mean magnetic field $\meanBB$ is determined by the equation
\EQ
\frac{\partial E_{\rm T}}{ \partial t} = I_{\rm T} - \frac{E_{\rm T}}{\tau_0} +
\etat ({\nab} \times \meanBB)^2  \;,
\label{energy1}
\EN
where $\tau_0$ is the  correlation time of the
turbulent velocity field in the maximum scale $\lf$ of
turbulent motions, $I_{\rm T}$ is the energy source of turbulence,
$\etat$ is the turbulent magnetic diffusion. For a given
time-independent source of turbulence $I_{\rm T}$ and for $t \gg \tau_0$
the total energy density of the turbulence reaches a steady state
\EQ
E_{\rm T} = \mbox{const} = \tau_0 \, I_{\rm T}\;,
\label{energy}
\EN
where we neglect a small magnetic source
$\etat ({\nab} \times \meanBB)^2$
of the turbulence [that is of the order of $O(\lf^2/H_B^2)$, and
$H_B$ is the characteristic scale of the mean magnetic
field spatial variations].
The approximate constancy of $E_{\rm T}$ is compatible with DNS,
where we found only a small decrease (20\%)
for strong (equipartition strength) mean fields
(see the upper panel of \Fig{pP64_pq}).
The reason for the departure is possibly a dependence of $\tau_0$ on $\meanB$.
However, this decrease only enhances the modification of the
Lorentz force by turbulence.

Equations~(\ref{eqstate}) and~(\ref{energy})
imply that the change of turbulent pressure $\delta \ptz$ is
proportional to the change of the magnetic  energy density
$\delta E_{\rm M}$, in particular $\delta \ptz = - (1 / 3)
\,\delta E_{\rm M}$
(because $\delta E_{\rm K} = -\delta E_{\rm M})$.
Therefore, the total turbulent pressure is
reduced when magnetic fluctuations are generated.

For a non-zero large-scale mean magnetic field $\meanBB$,
the change of the magnetic  energy density
$\delta E_{\rm M}$ is proportional to $\meanBB^2$. Therefore,
the total turbulent pressure is given by
$P_{\rm t} = \ptz - \half q_{\rm p} \, \meanBB^2$,
where $\ptz$ is the turbulent pressure
in a flow with a zero mean magnetic field.
The coefficient $q_{\rm p} > 0$ when magnetic
fluctuations are generated. Now we examine
the part, $P_{\rm eff}(\meanBB)$, of the total pressure, $P_{\rm tot} =
\overline{p} + P_{\rm t} + \half\meanBB^2$, that depends on the mean
magnetic field, $\meanBB$, i.e., we consider the
effective mean magnetic pressure that takes into account
the contribution of turbulence, $P_{\rm eff}(\meanBB)
= \half(1 - q_{\rm p}) \, \meanBB^2$. We study the case when
$\overline{p} \gg \meanBB^2/2$, and therefore, the total pressure
$P_{\rm tot}$ is always positive, while the effective
mean magnetic pressure $P_{\rm eff}(\meanBB)$ may
be negative when $q_{\rm p} >1$.

The modification of the mean Lorentz force can result in the excitation of
a long wavelength instability even in an initially uniform mean magnetic
field in a density stratified layer (Kleeorin et al.\ 1990, 1996;
Rogachevskii \& Kleeorin 2007).
Indeed, the growth rate of the instability for the perturbations
perpendicular to both the gravity field ${\bm g}$ and the mean magnetic
field $\meanBB$ is given by
\EQ
\lambda  = {c_{a} \over H_{\rho}} \sqrt{(1 - q_{\rm p}) \, \biggl({H_{\rho} \over H_{B}} - 1 \biggr)},
\label{growth-rate}
\EN
where $c_{a} = \meanB/\rho_0^{1/2}$ is the Alfv\'en speed,
$H_{\rho}^{-1}=|\nab\ln\rho_0|$, $H_{B}^{-1}=|\nab\ln\meanB|$,
and we neglected for simplicity the dissipation processes due to turbulent
viscosity and turbulent magnetic diffusion. For $q_{\rm p}>1$ the instability
is excited when $H_{\rho} < H_{B}$, i.e., it occurs even in an initially
uniform mean magnetic field.

The mechanism of the instability can be understood as follows.
An isolated magnetic tube moving upward becomes lighter than
the surrounding plasma since the decrease of the magnetic field in it,
due to expansion of the tube, is accompanied
with an increase of the magnetic pressure inside the tube.
This increase, due to the fact that the effective magnetic pressure
is negative, leads to a decrease of the density inside the tube and to
the appearance of a buoyancy force.
It results in the further upward displacement of the flux tube, i.e.\
it causes the excitation of the instability.
The instability causes the formation of inhomogeneous magnetic structures.
The energy for this instability is supplied by the small-scale turbulence.
In contrast, the free energy in Parker's magnetic buoyancy instability,
that is excited when $H_{\rho} > H_{B}$, is drawn from the gravitational
field (Newcomb 1961; Parker 1966).
The growth rate of Parker's magnetic buoyancy instability is
determined by \Eq{growth-rate} for $q_{\rm p} =0$.

Magnetic buoyancy in astrophysics applies usually to two different
situations (see, e.g., Priest 1982).
The first corresponds to a problem described by Parker (1966, 1979b)
and Gilman (1970a, 1970b).
They considered the instability of a stratified continuous magnetic field and
did not invoke a magnetic flux tube.
The other situation was considered by Parker (1955), Spruit (1981),
Spruit \& van Ballegooijen (1982), Ferriz-Mas \& Sch\"{u}ssler (1993),
and Sch\"{u}ssler et al.\ (1994), who studied buoyancy of horizontal
magnetic flux tubes.

In the present study we investigate the first situation, i.e., we study
the large-scale instability of a continuous (diffusive) magnetic field
in small-scale turbulence.
This instability is caused by turbulent velocity and magnetic fluctuations
and leads to the formation of large-scale magnetic structures (see Sect.\ 3).

\section{Mean-field numerical modelling}

In order to understand in more detail the appearance of magnetic
structures from this large-scale instability we consider now
numerical solutions of the
mean-field MHD equations in a density stratified layer.
We apply a new mean-field model which includes the effect of turbulence
on mean Lorentz force.
We solve the evolution equations for mean velocity $\meanUU$,
mean density $\meanrho$, and mean vector potential $\meanAA$ in the form
\EQ
{\partial\meanUU\over\partial t}=-\meanUU\cdot\nab\meanUU
-\cs^2\nab\ln\meanrho+\grav+\meanFFFF_{\rm M}+\meanFFFF_{\rm K,tot},
\EN
\EQ
{\partial\ln\meanrho\over\partial t}=-\meanUU\cdot\nab\ln\meanrho
-\nab\cdot\meanUU,
\EN
\EQ
{\partial\meanAA\over\partial t}=\meanUU\times\meanBB-(\etat+\eta)\meanJJ,
\EN
where $\meanFFFF_{\rm M}$ is given by \Eq{efforce}, and
$\meanFFFF_{\rm K,tot}=\meanFFFF_{\rm K}+\meanFF_{\rm visc}$
with $\meanrho\meanFFFF_{\rm K}=\nab\cdot\ssigma^{\rm K}$ and
with $\meanrho\meanFF_{\rm visc}=\nab\cdot\overline{\ssigma}$,
so that
\EQ
\meanFFFF_{\rm K,tot}=(\nut+\nu)\left(\nabla^2\meanUU+\nab\nab\cdot\meanUU
+2\meanSSSS\nab\ln\meanrho\right)
\EN
is the total (turbulent and microscopic) viscous force and
$\SSSS$ is given by \Eq{strain}.

\begin{figure}[t!]\begin{center}
\includegraphics[width=\columnwidth]{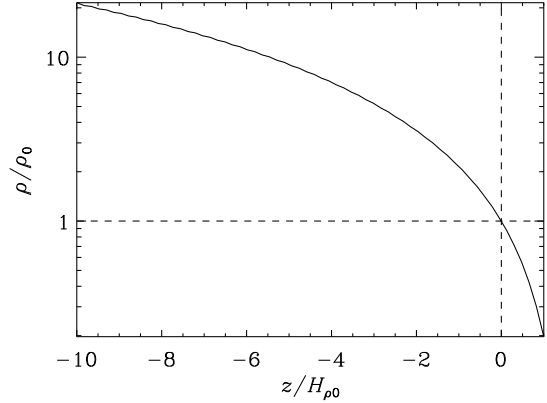}
\end{center}\caption[]{
Density stratification in a model with $z_{\rm top}=H_{\rho0}$.
The dashed lines indication the reference value with $\rho=\rho_0$ at $z=0$.
}\label{den}\end{figure}

We consider two- and three-dimensional models of an isentropically
stratified atmosphere, where the gravitational potential is written as
$\Phi(z)=(z-z_\infty)g$,
so the gravity vector $\grav=-\nab\Phi=(0,0,-g)$ is constant.
We arrange the initial profiles of density and sound speed such
that they take given reference values at $z=0$, i.e.\
$\meanrho=\rho_0$ and $\cs=\csz$ at $z=0$.
This implies that $z_\infty=(3/2)\csz^2/g$.
Our initial profiles therefore obey
\EQ
\rho/\rho_0=(\cs/\csz)^3,\quad
\cs^2=-{\textstyle{2\over3}}\Phi.
\EN
The local density scale height is $H_\rho=\cs^2/g$, and the
density scale height at $z=0$ is $H_{\rho0}=\csz^2/g$.
In \Fig{den} we show the vertical dependence of the initial density.

We allow for the presence of an imposed field
in the $y$ direction, $\BB_0=(0,B_0,0)$.
The total field is then written as
\EQ
\meanBB=\BB_0+\nab\times\meanAA,
\EN
so the departure from the imposed field is expressed in terms
of the mean magnetic vector potential $\meanAA$.

On the upper and lower boundaries
we adopt stress-free boundary conditions for velocity,
i.e.\ $\meanU_{x,z}=\meanU_{y,z}=\meanU_{z}=0$, and a normal-field
condition for the magnetic field, i.e.\ $\meanB_x=\meanB_y-B_0=0$,
corresponding to $\meanA_{x,z}=\meanA_{y,z}=\meanA_{z}=0$
for the vector potential.
Here, commas denote partial differentiation.
No boundary condition for the density is required.
Again, all computations have been carried out with the {\sc Pencil Code}.

In setting up our model we define the wavenumber $\kf$ of the
energy-carrying eddies.
This relates the turbulent magnetic diffusivity to the
rms velocity via $\etat=\urms/3\kf$.
This means that the ratio of our non-dimensional field strength
to turbulent diffusivity, i.e.\ $B_0/\csz\rho_0^{1/2}$
to $\etat/\csz H_\rho$, is given by $3\kf H_0$ times $B_0/B_{\rm eq}$.

\begin{figure}[t!]\begin{center}
\includegraphics[width=\columnwidth]{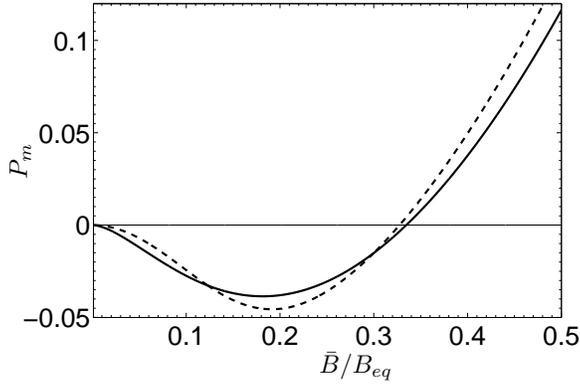}
\end{center}\caption[]{
The effective mean magnetic pressure $P_m({\meanB}) = (1-q_{\rm p}) \meanB^2
/ \meanB_{\rm p}^2$ determined by Rogachevskii \& Kleeorin (2007) -- solid
line, and by the model described by Eq.~(\ref{qp}) -- dashed line
($\meanB_{\rm p}=0.21\,\csz\rho_0^{1/2}$ and $q_{{\rm p}0}=4$).
}\label{pstrat}\end{figure}

\begin{figure}[t!]\begin{center}
\includegraphics[width=\columnwidth]{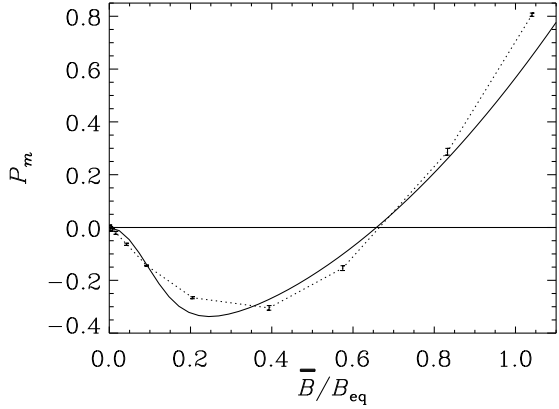}
\end{center}\caption[]{
Same as \Fig{pstrat}, but from simulation (dotted line).
The solid line shows a fit [Eq.~(\ref{qp})] with
$\meanB_{\rm p} = 0.022 \, \csz\rho_0^{1/2}$ (corresponding to
$\meanB_{\rm p}/B_{\rm eq}=0.18$) and $q_{{\rm p}0}=21$.
}\label{pP64_press}\end{figure}

In this paper we approximate $q_{\rm p}$ and $q_{\rm s}$ by simple profile functions,
\EQ
q_{\rm p}=q_{{\rm p}0}\left(1-{2\over\pi}\arctan{\meanBB^2\over
\meanB_{\rm p}^2}\right),
\label{qp}
\EN
\EQ
q_{\rm s}=q_{{\rm s}0}\left(1-{2\over\pi}\arctan{\meanBB^2\over
\meanB_{\rm s}^2}\right).
\label{qs}
\EN
Following Rogachevskii \& Kleeorin (2007),
we also define $Q_{\rm p}=1-q_{\rm p}$ and $Q_{\rm s}=1-q_{\rm s}$.
Correspondingly, we define the coefficients
$Q_{{\rm p}0}=1-q_{{\rm p}0}$ and $Q_{{\rm s}0}=1-q_{{\rm s}0}$.
For our fiducial model we use $Q_{{\rm p}0}=2Q_{{\rm s}0}=20$,
corresponding to $q_{{\rm p}0}=21$ and $q_{{\rm s}0}=11$.
The resulting magnetic pressure is shown in \Fig{pstrat} for their analytic
theory and the result from DNS is shown in \Fig{pP64_press}, together with
the corresponding fits.
For our fiducial model we choose furthermore
$\meanB_{\rm p}=\meanB_{\rm s}=0.1\,\csz\rho_0^{1/2}$.
For the imposed field strength we choose
$B_0/\csz\rho_0^{1/2}=0.01$ and for the turbulent magnetic
diffusivity we take $\etat/\csz H_\rho=0.01$.
As discussed above, this means that
$B_0/B_{\rm eq}=1/3$ if we assume $\kf H_0=1$.

\subsection{Magnetic structures in two-dimensions}
\label{MagneticStructures}

In this paper we consider both two-dimensional and three-dimensional
solutions.
We begin with two-dimensional models in the $xz$ plane with an imposed
field in the normal ($y$) direction, $\BB_0=(0,B_0,0)$.

\begin{table}[bt]\caption{
Summary of non-dimensional run parameters together with the resulting
non-dimensional growth rates $\tilde\lambda$ as well as the non-dimensional
saturation values of rms velocity and magnetic field.
The tildes indicate non-dimensional quantities, as explained in
\Sec{MagneticStructures}.
Our fiducial run is Run~C.
}\vspace{12pt}\centerline{\begin{tabular}{lcccccccc}
&
$\!\!\tilde{B}_0\!\!$ &
$\!\!\tilde\etat\!\!$ &
$\!\!\!\!\!\!\tilde{z}_{\rm top}\!\!\!\!\!\!$ &
$\!\!Q_{p0}\!\!\!\!$ &
$\!\!Q_{s0}\!\!\!\!$ &
$\!\!\tilde\lambda\!\!$ &
$\!\!\tilde{U}_{\rm sat}\!\!$ &
$\!\!\tilde{B}_{\rm sat}\!\!$ \\
\hline
A & 0.01 & 0.01 & 1 &  $-4$ &  $-2$ &$\!\!0.0005\!\!$& 0.000 & 0.000 \\
B & 0.01 & 0.01 &$\!\!1.2\!\!$& $-20$ & $-10$ & 0.012 & 0.013 & 0.013 \\ 
C & 0.01 & 0.01 & 1 & $-20$ & $-10$ & 0.006 & 0.013 & 0.013 \\ 
D & 0.02 & 0.01 & 1 & $-20$ & $-10$ & 0.032 & 0.025 & 0.026 \\ 
E & 0.01 & 0.02 & 1 & $-20$ & $-10$ & 0.001 & 0.001 & 0.010 \\ 
\\
\\
\label{Summary}\end{tabular}}\end{table}

In \Tab{Summary} we present a summary of some exploratory runs where we
list the nondimensional growth rate $\tilde\lambda\equiv\lambda H_\rho^2/\etat$,
as well as the nondimensional saturation values of rms mean velocity and
mean magnetic field, $\tilde{U}_{\rm rms}\equiv\meanU_{\rm rms}H_{\rho0}/\etat$
and $\tilde{B}_{\rm rms}\equiv\meanB_{\rm rms}H_{\rho0}/\etat\rho_0^{1/2}$.
These experiments show that decreasing $Q_{{\rm p}0}$ and $Q_{{\rm s}0}$ lowers
the growth rate and the saturation values, while increasing the degree
of stratification (e.g., increasing
$\tilde{z}_{\rm top}\equiv z_{\rm top}/H_{\rho0}$
from 1 to 1.2, corresponding to an increase of bottom to top density ratio
from 108 to 237) enhances them.
Likewise, increasing the strength of the imposed field enhances the
growth rate and the saturation values, while increasing the magnetic
diffusivity lowers them.
These results are in agreement with \Eq{growth-rate}.

\begin{figure}[t!]\begin{center}
\includegraphics[width=\columnwidth]{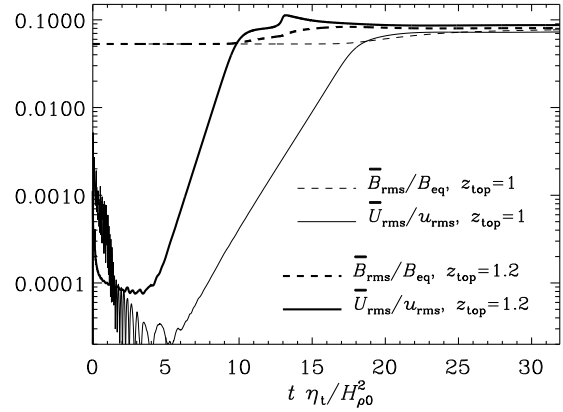}
\end{center}\caption[]{
Growth of the rms value of mean velocity and mean magnetic field
for two runs with different degree of stratification.
}\label{pn_comp_strat}\end{figure}

\begin{figure}[t!]\begin{center}
\includegraphics[width=\columnwidth]{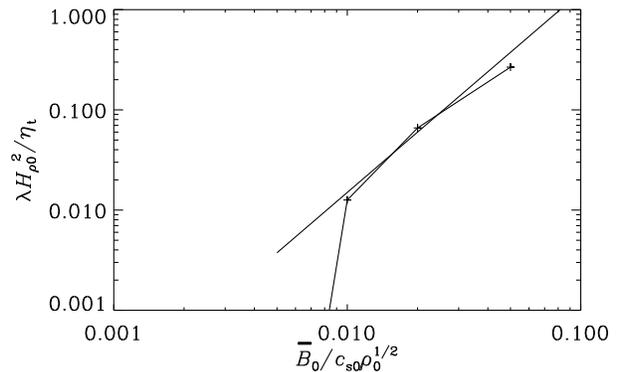}
\end{center}\caption[]{
Growth rate as a function of $B_0$, keeping all other parameters
as for the fiducial Run C.
}\label{plam_vs_b0}\end{figure}

In \Fig{pn_comp_strat} we compare the evolution of the rms values of
velocity and magnetic field for two different stratifications.
As discussed above, increasing the amount of stratification increases the
growth rate.
The scaling of the growth rate with the strength of the imposed field
is shown in \Fig{plam_vs_b0}.

\begin{figure}[t!]\begin{center}
\includegraphics[width=\columnwidth]{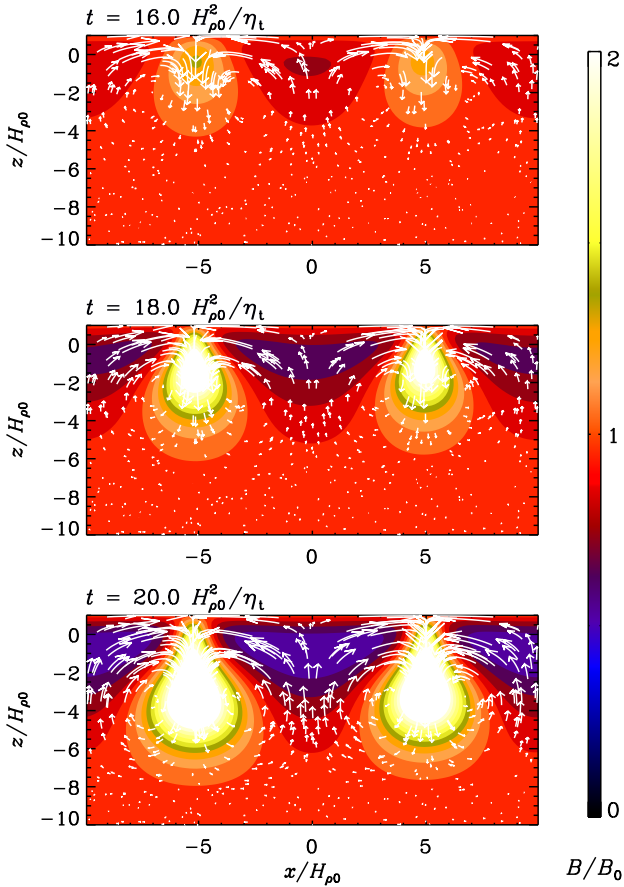}
\end{center}\caption[]{
Early evolution of magnetic field in the $y$ direction (color coded)
together with velocity vectors in the $xz$ plane.
Note the spontaneous production of flux structures.
}\label{pbcomp_64x64_strat1_B1_W1a}\end{figure}

Next, we consider the structure of the resulting magnetic field.
In \Fig{pbcomp_64x64_strat1_B1_W1a} we show ``meridional'' ($xz$)
cross-sections of the magnetic field at three different times during
the early phase where the magnetic field just begins to saturate.
The horizontal wavelength of the pattern is about $10\,H_\rho$.
As time goes on, the structures of enhanced magnetic field intensify.
The decrease of the effective pressure makes them even heavier
which explains their subsequent descent.

\begin{figure}[t!]\begin{center}
\includegraphics[width=\columnwidth]{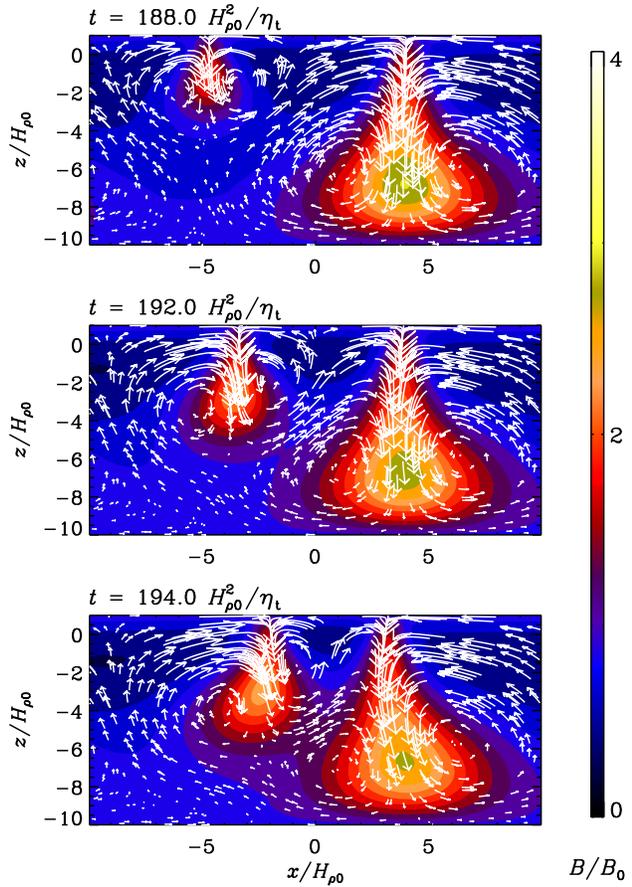}
\end{center}\caption[]{
Later evolution of magnetic field for the same run as in
\Fig{pbcomp_64x64_strat1_B1_W1a}.
Note the mutual merging of flux structures.
}\label{pbcomp_64x64_strat1_B1_W1b}\end{figure}

At later times new structures can form near the surface.
In \Fig{pbcomp_64x64_strat1_B1_W1b} we show an example during the
fully saturated phase of the instability, where one sees the
emergence of a new patch that is then swept to the larger one
and merges.

\subsection{Magnetic structures in three-dimensions}

\begin{figure}[t!]\begin{center}
\includegraphics[width=\columnwidth]{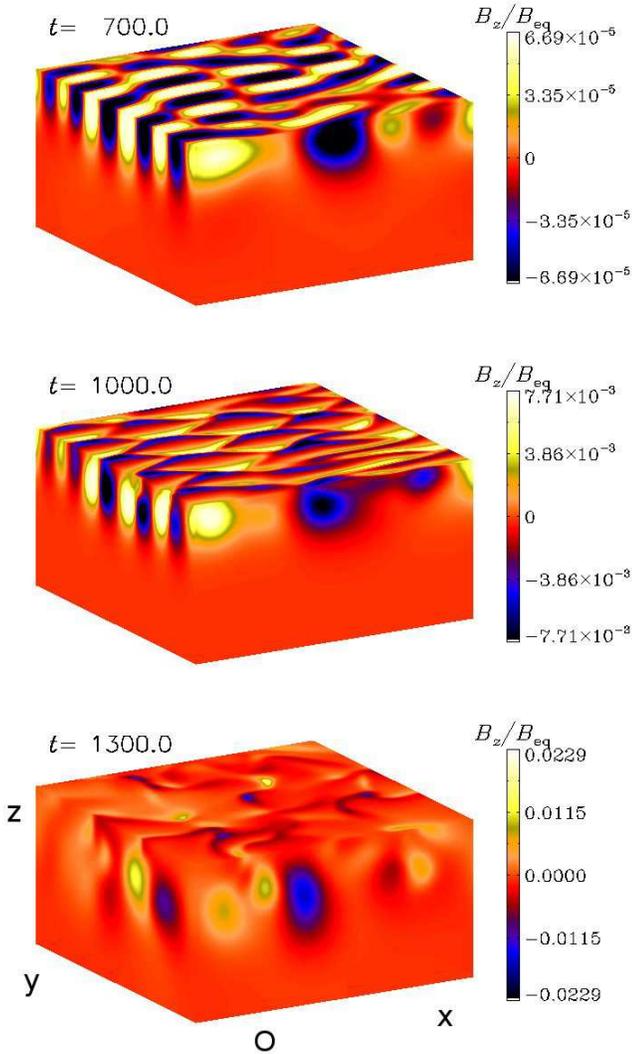}
\end{center}\caption[]{
Visualizations of the magnetic field at the
early saturation phase ($t=700\,H_{\rho0}/\csz=7\,H_{\rho0}^2/\etat$ left),
at an intermediate time ($t=1000\,H_{\rho0}/\csz=10\,H_{\rho0}^2/\etat$ left),
and a later time ($t=1300\,H_{\rho0}/\csz=13\,H_{\rho0}^2/\etat$ left).
Note the broad similarity of field in the $xz$ plane with the
two-dimensional cases.
Note that the wavelength of the pattern is shorter in the
$y$ direction than in the $x$ direction.
In the final time much of the magnetic field structures have sunk beneath
the surface, leaving only a few isolated bipolar structures at the surface.
}\label{64x64_strat1_B1_W1r_3D}\end{figure}

In order to demonstrate the three-dimensional nature of the instability
we extend the domain in the $y$ direction, so both $x$ and $y$
are between $\pm10H_{\rho0}$, and $-10\leq z/H_{\rho0}\leq1$.
These simulations are otherwise similar to those in the two-dimensional cases.
In \Fig{64x64_strat1_B1_W1r_3D} we show visualizations of the field
at three characteristic times.
Note in particular that the wavelength of the pattern in the
$y$ direction (parallel to the field) is 3--4 times shorter than
that in the $x$ direction (perpendicular to the field).
Again, the instability begins to emerge first near the surface
where it develops magnetic structures which begin to sink downwards.
Viewed from above, one sees the emergence of what looks like
multiple bipolar regions.

\begin{figure}[t!]\begin{center}
\includegraphics[width=\columnwidth]{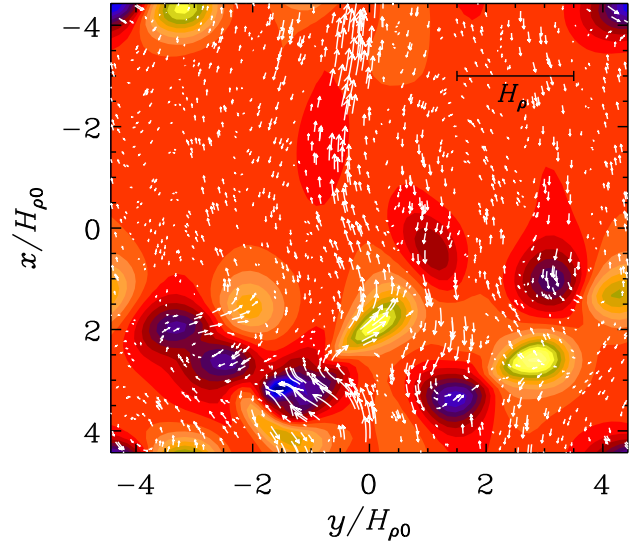}
\end{center}\caption[]{
Magnetic field in the $xy$ plane.
Note that the plane has been rotated by $90^\circ$ in the counterclockwise
direction, so the $y$ direction, corresponding to the azimuthal
direction when applied to the Sun, points from left to right.
The black horizontal bar gives the density scale height at
the depth of the cross-section, which is about twice the
value $H_{\rho0}$ at the reference depth.
}\label{pbxy}\end{figure}

In \Fig{pbxy} we show a horizontal cross-section from another simulation,
where the horizontal extend is only half as much as before.
This figure is suggestive of the formation of bipolar structures.
The plane has been rotated by $90^\circ$ such that the $y$ direction
points now from left to right.
The black horizontal bar gives the density scale height at
the depth of the cross-section, which is about twice the
value $H_{\rho0}$ at the reference depth.

\section{Conclusions}

In this study we have demonstrated in DNS the effects
of turbulence on the mean Lorentz force. This effect
is quantified by determining the relevant functions $q_{\rm p}(\meanB)$
and $q_{\rm s}(\meanB)$ that relate the sum of the turbulent Reynolds
and Maxwell stresses with the Maxwell stress of the mean magnetic field.
Using three-dimensional simulations of forced hydromagnetic
turbulence with an imposed field,
we confirm that the function $q_{\rm p}(\meanB)$ is positive
and can reach values much larger than unity for $\meanB/\Beq\ll1$.
This thereby reverses the sign of the effective
magnetic pressure $P_{\rm eff}(\meanBB)$ associated with
the mean magnetic field, which then becomes $P_{\rm eff}(\meanBB)=
\half(1-q_{\rm p})\meanB^2$. We find that the function $q_{\rm s}(\meanBB)$
that determines the modification of magnetic tension,
also reaches values much larger than unity, but its value is less
than half the value of $q_{\rm p}$ and the error bars are larger.
This work has also demonstrated explicitly the possibility of a large-scale
instability of the full system of mean-field MHD equations.
Finally, our study has revealed the presence of spatial structures
arising from the instability that might be associated with the formation
of bipolar magnetic regions and perhaps also sunspots when applied to the Sun.

The effects of turbulence on the large-scale Lorentz force
may also be important in applications to the solar torsional oscillations
by changing the mean magnetic tension $\meanBB \cdot \nab \meanBB$
(R\"{u}diger et al.\ 1986; R\"udiger \& Kichatinov 1990).
More specifically, these turbulence effects may be critical for
explaining the narrow structure of the
observed solar torsional oscillations (Kleeorin et al.\ 1996).

Clearly, there are several possibilities of improvement that might
make the model more realistic and eventually suitable for
confrontation against observations.
Most important is perhaps the fact that our reference values $B_{{\rm p}0}$
and $B_{{\rm s}0}$ in the quenching profiles \eqs{qp}{qs}, as well as the
coefficients $q_{{\rm p}0}$ and $q_{{\rm s}0}$ in these profile functions, are
currently kept constant.
It will be more realistic to make them vary with depth, because density
and turbulent rms velocity vary with depth.
Another extension of the model would be to allow for the
possibility that the magnetic field to be generated by
a mean-field dynamo rather than relying on an imposed field.

On the more theoretical side, there is a need to further verification
of the essential physics of the negative magnetic pressure effect.
In particular, one must wonder why the effects of this instability have
not yet seen in DNS.
A likely possibility is the lack of sufficient scale separation.
Only now realistic simulations are beginning to be large
enough to encompass sufficiently many cells in all three directions.
The importance of sufficient horizontal extent was already emphasized
by Tao et al.\ (1998), who find clear evidence of a segregation into
strongly magnetized and weak magnetized regions, a phenomenon that
might be closely related to that reported here.
A particularly promising approach might be to generalize the
direct simulations discussed in the present paper to the case with
vertical density stratification such that the setup becomes similar
to the mean field models that we also studied in this paper.
The turbulence would then still be driven by a forcing function.
Another possibility is to study this effect with turbulent convection
instead of forced turbulence.
This is particularly interesting, because theoretical predictions
of Rogachevskii \& Kleeorin (2007) suggest that the modification
of the effective Lorentz force will be even stronger in
turbulent convection.

\section*{Acknowledgments}
We have benefited from very useful discussions with Alexander Kosovichev.
We acknowledge the use of computing time at the Center for
Parallel Computers at the Royal Institute of Technology in Sweden
as well as the National Supercomputing Center in Link\"oping.
This work was supported in part by the European Research Council under
the AstroDyn Research Project 227952 and the Swedish Research Council
grant 621-2007-4064.
NK and IR thank NORDITA for hospitality and support during their visits.
The final stage of this work was completed while participating at
the NORDITA program on ``Solar and stellar dynamos and cycles''.

\end{document}